\documentclass[%
 aip,
 jmp,%
 amsmath,amssymb,
 reprint,%
]{revtex4-1}

\usepackage{graphicx}% Include figure files
\usepackage{dcolumn}% Align table columns on decimal point
\usepackage{bm}% bold math
\begin{document}

\preprint{AIP/123-QED}

\title[Trapping in dendrimers and regular hyperbranched polymers]{Trapping in dendrimers and regular
hyperbranched polymers}% Force line breaks with \\
%\thanks{Footnote to title of article.}

\author{Bin Wu}
\author{Yuan Lin}
\author{Zhongzhi Zhang}
\email{zhangzz@fudan.edu.cn}

\affiliation {School of Computer Science, Fudan University,
Shanghai 200433, China}

\affiliation {Shanghai Key Lab of Intelligent Information
Processing, Fudan University, Shanghai 200433, China}

\author{Guanrong Chen}
\affiliation {Department of Electronic Engineering, City University of Hong Kong, Hong Kong SAR, China}

\date{\today}% It is always \today, today,

\begin{abstract}
Dendrimers and regular hyperbranched polymers are two classic families of macromolecules, which can be modeled by Cayley trees and Vicsek fractals, respectively. In this paper, we study the trapping problem in Cayley trees and Vicsek fractals with different underlying geometries, focusing on a particular case with a perfect trap located at the central node. For both networks, we derive the exact analytic formulas in terms of the network size for the average trapping time (ATT)---the average of node-to-trap mean first-passage time over the whole networks. The obtained closed-form solutions show  that for both Cayley trees and Vicsek fractals, the ATT display quite different scalings with various system sizes, which  implies that the underlying structure plays a key role on the efficiency of trapping in polymer networks. Moreover, the dissimilar scalings of ATT may allow to differentiate readily between dendrimers and hyperbranched polymers.
\end{abstract}

%05.40.Fb Random walks and Levy flights
%61.43.Hv  Fractals; macroscopic aggregates (including diffusion-limited aggregates)
%05.45.Df Fractals
%05.60.Cd Classical transport
%05.40.-a Fluctuation phenomena, random processes, noise, and Brownian motion
%89.75.Hc Networks and genealogical trees

\maketitle

%\tableofcontents

\section{Introduction}

In the last few decades, polymer physics has attracted considerable attention within the scientific community,
with various polymer networks proposed to describe the structures of macromolecules~\cite{GuBl05}.  Among
numerous polymer networks, Cayley trees and Vicsek fractals are two important ones, both of which have a
treelike structure for modeling dendrimers and regular hyperbranched macromolecules, respectively.  The
treelike dendrimers consist of repeating units arranged in a hierarchical, self-similar way around a central
core~\cite{ToNaGo90,ArAs95}. These special properties make them promising candidates for a large number of
applications, e.g., light harvesting antennae~\cite{BaCaDeAlSeVe95,KoShShTaXuMoBaKl97}.  Because of their
practical significance, much attention has been devoted to the investigation of
dendrimers~\cite{CaCh97,ChCa99,SuHaFrMu99,SuHaFr00,GaFeRa01,BiKaBl01,MuBiBl06,GaBl07,MuBl11}.

Despite that dendrimers are of theoretical and practical interest, from the view point of chemistry, they are
not simple to prepare~\cite{ToNaGo90}. Thus, another class of polymers without this deficiency is desirable,
which are hyperbranched polymers that are much easier to synthesize~\cite{SuHaFrMu99,SuHaFr00}.  Some
hyperbranched polymers may be regular fractals, a particular example of which is the classic Vicsek fractals,
which were first introduced in~\cite{Vi83} and were extended in~\cite{BlJuKoFe03,BlFeJuKo04}. As one of the
most important regular fractals, Vicsek fractals have attracted extensive
interest~\cite{Vi84,WaLi92,JaWu92,JaWu94,StFeBl05,ZhZhChYiGu08,Vo09,ZhWjZhZhGuWa10,JuvoBe11} and continue to
be an active object of research in various areas~\cite{Fa03,AgViSa09}.

As is well known, a fundamental topic in polymer physics is to reveal how the underlying topologies of
polymeric materials influence their dynamic behavior~\cite{GuBl05}.  Among plethora dynamical processes,
trapping is a paradigmatic one, which is a kind of random walk with a deep trap fixed at a given position,
absorbing all walkers that visit it. Many dynamical processes in macromolecular systems can be described as a
trapping process, e.g., lighting harvesting~\cite{BaKlKo97,BaKl98}. A basic quantity relevant to the trapping
problem is the trapping time (TT), commonly called the mean first-passage time (MFPT), for general random
walks~\cite{Re01,Lo96,MeKl04,BuCa05}. The TT for a node $i$, denoted by $F_i$, is the expected time for a walker
starting from $i$ to reach the trap for the first time. The average trapping time (ATT), $\langle F \rangle$,
is defined as the average of $F_i$ over all source nodes in the system other than the trap, which provides a
useful indicator for the efficiency of  trapping.  Thus far, trapping problem has been extensively studied for
various complex systems, such as regular lattices~\cite{Mo69}, the Sierpinski
gasket~\cite{KaBa02PRE,KaBa02IJBC}, the $T-$fractal~\cite{KaRe89,Ag08,HaRo08,LiWuZh10,ZhWuCh11}, as well as various
scale-free
graphs~\cite{KiCaHaAr08,ZhQiZhXiGu09,ZhZhXiChLiGu09,AgBu09,ZhLiGoZhGuLi09,TeBeVo09,AgBuMa10,ZhYaLi12,MeAgBeRo12}.
However, trapping problem for Cayley trees and Vicsek fractals is still not well understood, in spite of that
they well describe these two important classes of polymers.

In this paper, we study analytically the trapping issue in Cayley trees and Vicsek fractals, which  are
typical polymer networks. Their special structures make them promising candidates as artificial antennae, with
their centers being the fluorescent traps. We thus focus on a special case of the trapping problem with the
trap placed at the central node. We will determine  closed-form formulae of ATT for both polymer networks, by
taking the advantage of the specific constructions of the polymer systems. The obtained explicit expressions
indicate that for very large systems the dominating scalings of ATT for the two systems display distinct
behaviors with respect to the system sizes. Our work sheds some lights on the concerned trapping problem,
providing some relevant relation information between trapping efficiency and underlying geometry of the
system.

\section{Introduction to Cayley trees and Vicsek fractals}

Here, we introduce the constructions and some properties of Cayley trees and Vicsek fractals
as two representative models of polymer networks. Both networks
are defined in an iterative way. Their particular
constructions allow for precisely analyzing their properties and  obtaining explicit closed-form
solutions for various dynamical processes on large but finite structures.

\subsection{Cayley trees}

\begin{figure}
\begin{center}
%\fbox{\includegraphics[width=.85\linewidth,trim=100 0 100 0]{Vicsek}}
\includegraphics[width=0.93\linewidth,trim=0 0 0 0]{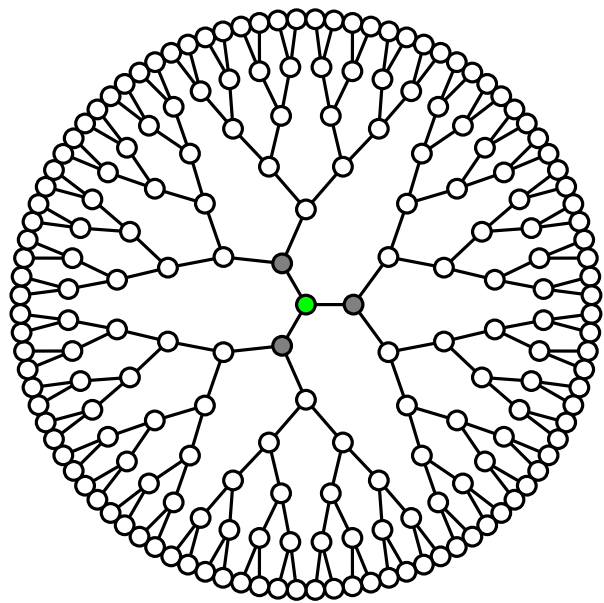}
\caption{(Color online) The Cayley tree $C_{3,6}$. The filled circles represent $C_{3,1}$.}  \label{Cayley}
\end{center}
\end{figure}

Let  $C_{m,g}$ ($m\geq 3$, $g \geq 0$) denote the Cayley trees after $g$ iterations (generations), which
can be constructed  as follows. At the initial generation ($g=0$), $C_{m,0}$ contains only a central  node
(the core); at $g=1$,  $m$ nodes are created attaching the central node to form $C_{m,1}$, with the $m$
single-degree nodes  constituting the boundary nodes of $C_{m,1}$. For any $g>1$,  $C_{m,g}$ is obtained
from $C_{m,g-1}$: For each peripheral node of $C_{m,g-1}$, $m-1$ new nodes are generated and are linked to the peripheral node.
Figure~\ref{Cayley} shows a particular Cayley tree, $C_{3,6}$. Let $N_i(g)$ be the number of nodes in
$C_{m,g}$, which are born in generation  $i$. Then, it is easy to verify that
\begin{equation}\label{cay1}
N_i(g)=\begin{cases}
1, &i=0,\\
m(m-1)^{i-1}, &i>0.\\
\end{cases}
\end{equation}
Thus, the total number of nodes in $C_{m,g}$ is
\begin{equation}\label{cay1a}
N_g=\sum_{i=0}^{g}N_{i}(g)=\frac{m(m-1)^g-2}{m-2}\,.
\end{equation}
Note that Cayley trees are nonfractal objects, irrespective of their self-similar structures; that is, their
fractal
dimension is infinite.

%%%%%%%%%%%%%%%%%%%%%%%%%%%%%%%%%%%%%%%%%%%%%%%%%%%%%%%%%
% Figure  1
%%%%%%%%%%%%%%%%%%%%%%%%%%%%%%%%%%%%%%%%%%%%%%%%%%%%%%%%%%
%%%%%%%%%%%%%%%%%%%%%%%%%%%%%%%%%%%%%%%%%%%%%%%%%%%%%%%%%%

\subsection{Vicsek fractals}

\begin{figure}
\begin{center}
%\fbox{\includegraphics[width=.85\linewidth,trim=100 0 100 0]{Vicsek}}
\includegraphics[width=1.02\linewidth,trim=0 0 0 0]{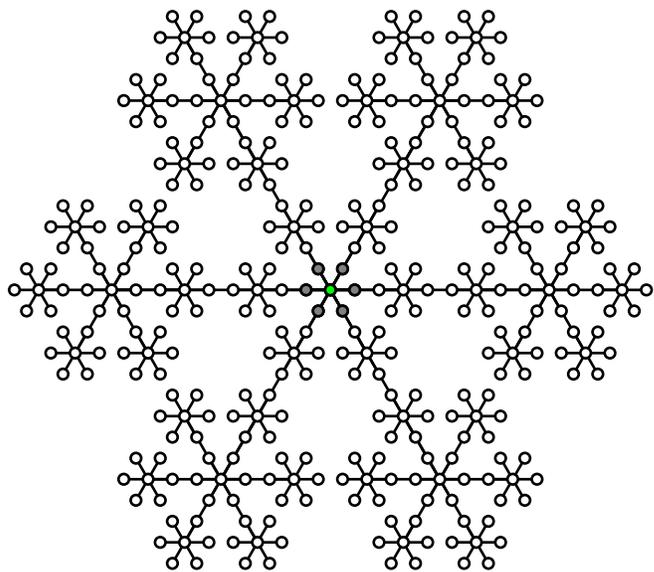}
\caption{(Color online) Illustration of the first several iterative processes of a
special Vicsek fractal, $V_{6,3}$. The filled circles denote the
starting structure $V_{6,1}$.} \label{Vicsek}
\end{center}
\end{figure}

As another new class of polymer networks, the Vicsek fractals are constructed in a different iterative way
~\cite{Vi83,BlJuKoFe03}. Let $V_{f,g}$ ($f\geq 2$, $g \geq 1$) denote the Vicsek fractals
after $g$ iterations (generations). For $g=1$, $V_{f,1}$ is a star-like cluster consisting of
$f+1$ nodes arranged in a cross-wise pattern, where a central node is connected to $f$ peripheral nodes.  For
$g\geq 2$, $V_{f,g}$ is obtained from $V_{f, g-1}$.
To obtain $V_{f,2}$, first $f$ replicas of $V_{f,1}$ are generated,  and then
arranged  around the periphery of the original $V_{f,1}$. They are connected to the central structure by $f$
additional links. These replication and connection steps are repeated
infinitely many times, with the  Vicsek fractals obtained in the limit $g
\rightarrow \infty$. In Fig.~\ref{Vicsek}, we show schematically the structure of $V_{6,3}$.
According to the above construction algorithm, at each step the number of
nodes in the system increases by a factor of $f+1$; thus the total number of nodes  of $V_{f,g}$ is $N_{g}=
(f+1)^{g}$. Since the whole
family of Vicsek fractals has a treelike structure, the total number
of links in $V_{f,g}$ is $E_{g}= N_{g}-1=(f+1)^{g}-1$.

%%%%%%%%%%%%%%%%%%%%%%%%%%%%%%%%%%%%%%%%%%%%%%%%%%%%%%%%%
% Figure  2
%%%%%%%%%%%%%%%%%%%%%%%%%%%%%%%%%%%%%%%%%%%%%%%%%%%%%%%%%%
%%%%%%%%%%%%%%%%%%%%%%%%%%%%%%%%%%%%%%%%%%%%%%%%%%%%%%%%%%

Differing from the Cayley  trees, Vicsek fractals are fractal objects just as their name suggests, with the
fractal
dimension being  $\ln (f+1)/\ln3$.

\section{Trapping with a single trap at the central node}

After introducing the two polymer networks, Cayley trees and Vicsek fractals, in this
section we study a particular random walk---the trapping problem---performed on  $C_{m,g}$ and  $V_{f,g}$,
where a single immobile trap is located at the central node.

The random-walk model considered here is a simple one. At
each discrete time step, the walker (particle) jumps from its current position to any of its neighboring nodes with an
identical probability. For convenience, the central node of
$C_{m,g}$ (or  $V_{f,g}$) is  labeled by $1$ , while all other nodes are labeled consecutively as $2$, $3$,
$\ldots$, $N_{g}-1$, and $N_{g}$. Let $F_{i}(g)$ denote the trapping time for node $i$, which is the expected
time for a walker staring from node $i$ to first
arrive at the trap in $C_{m,g}$ (or  $V_{f,g}$). The most important
quantity related to the trapping problem is the ATT,  $\langle F \rangle_{g}$, which is the average of
$F_{i}(g)$ over all starting nodes distributed uniformly over the whole network. By
definition, $\langle F \rangle_{g}$ is given by
\begin{equation}\label{ATT}
\langle F \rangle_{g}=\frac{1}{N_g-1}\sum_{i=2}^{N_g} F_i(g)\,.
\end{equation}

In the sequel, we will determine explicitly this $\langle
F \rangle_{g}$ for both $C_{m,g}$ and  $V_{f,g}$, and show how $\langle F \rangle_{g}$ scales with  the system size, so as to get information on the internal structure of the polymer networks and explore the effect of the underlying geometry on the trapping efficiency on the networks.

\subsection{Average trapping time for Cayley trees}

Note that all nodes in $C_{m,g}$ can be classified into $g+1$ levels. The central node is at level $0$, the
nodes created at generation $1$ are at level $1$, and so on.
By symmetry, all nodes at the same level have the same TT. In the case without confusion,  $F_{i}(g)$ is used to
represent the TT  for a node at level $i$ in $C_{m,g}$,  which satisfies the following relations:
\begin{equation}\label{Cayley01}
F_i(g)=\begin{cases}
0, &i=0,\\
\frac{1}{m}[1+F_{i-1}(g)]+\frac{m-1}{m}[1+F_{i+1}(g)],&0<i<g,\\
1+F_{g-1}(g), &i=g\,.\\
\end{cases}
\end{equation}
For $i=0$ and $i=g$, Eq.~(\ref{Cayley01}) is obvious; while for $0<i<g$, it can be elaborated as follows. The first term on the right-hand side accounts for the case that with probability $\frac{1}{m}$ the walker starting from a node at level $i$ first takes one time step
to arrive at its unique neighbor at level $i-1$ and then takes $F_{i-1}(g)$ steps to reach the trap for the first time. The second term explains the fact that with probability $\frac{m-1}{m}$ the walker fist makes a jump to a node at level $i+1$ and then jumps $F_{i+1}(g)$ more steps to first reach the central node.

Thus, for $0<i<g$, one has
\begin{equation}\label{Cayley02}
F_{i}(g)-F_{i-1}(g)=m+(m-1)[F_{i+1}(g)-F_{i}(g)]\,.
\end{equation}
Let $A_{i}(g)=F_{g-i}(g)-F_{g-i-1}(g)$. Then,
\begin{equation}\label{Cayley03}
A_{i}(g)=m+(m-1) A_{i-1}(g)\,
\end{equation}
holds for all $0<i<g$.
Using the initial condition $A_{0}(g)=T_{g}(g)-T_{g-1}(g)=1 $,   Eq.~(\ref{Cayley03}) can be solved to yield
\begin{equation}\label{Cayley04}
A_{i}(g)=\frac{1}{m-2}[2(m-1)^{i+1}-m]\,.
\end{equation}
That is, for $i<g$, one has
\begin{equation}\label{Cayley05}
F_{g-i}(g)=\frac{1}{m-2}[2(m-1)^{i+1}-m]+F_{g-i-1}(g)\,,
\end{equation}
which leads to
\begin{equation}\label{Cayley06}
F_{i}(g)=\frac{1}{m-2}[2(m-1)^{g-i+1}-m]+F_{i-1}(g)\,,
\end{equation}
for all $i>0$.
Since  $F_{0}(g)=0 $, Eq.~(\ref{Cayley06}) can be solved to yield
\begin{equation}\label{Cayley07}
F_{i}(g)=\frac{2}{(m-2)^{2}}\left[(m-1)^{g+1}-(m-1)^{g-i+1}\right]-\frac{m}{m-2} i\,,
\end{equation}
for all $i>0$.

Then, according to Eq.~(\ref{ATT}), the explicit expression for ATT for the trapping problem in $C_{m,g}$ can be
obtained as
\begin{eqnarray}\label{Cayley08}
\langle F \rangle_{g}&=&\frac{\sum_{i=1}^{g}[N_{i}(g)\times F_{i}(g)]}{N_g-1}\nonumber\\
&=&\frac{2(m-1)^{2g+1}-(m-2)(m-1)^g[(m+2)g+1]-m}{(m-2)^{2}[(m-1)^{g}-1]}\,.\nonumber\\
\end{eqnarray}
We proceed to  represent $\langle F \rangle_g$ as
a function of the system size $N_g$. From Eq.~(\ref{cay1a}), we have
\begin{equation}\label{Cayley09}
g=\frac{\ln [(m-2)N_g+2] - \ln m}{\ln (m-1)}\,
\end{equation}
which enables to write
$\langle F \rangle_g$ in the following form:
\begin{eqnarray}\label{Cayley10}
&\quad&\langle F \rangle_{g}\nonumber\\
&=&\frac{2m-2}{m^2-2m}\frac{(N_g)^2}{N_g-1}\nonumber\\
&&-\frac{m+2}{(m-2)\ln(m-1)}\frac{N_g\ln [(m-2)N_g+2]}{N_g-1}\nonumber\\
&&-\left[\frac{m^2-10m+8}{m(m-2)^2}-\frac{(m+2)\ln m}{(m-2)\ln(m-1)}\right]\frac{N_g}{N_g-1}\nonumber\\
&&-\frac{2m+4}{(m-2)^2\ln(m-1)}\frac{\ln [(m-2)N_g+2]}{N_g-1}\nonumber\\
&&-\left[\frac{m^2+4m-4}{m(m-2)^2}-\frac{(2m+4)\ln m}{(m-2)^2 \ln (m-1)}\right]\frac{1}{N_g-1}\,. \nonumber\\
\end{eqnarray}
Equation~(\ref{Cayley10}) provides the exact dependence relation
of ATT on the network size $N_g$ and parameter $m$. For a large
system, i.e., $N_g\rightarrow \infty$, we have the following expression
for the dominating term of $\langle F \rangle_g$:
\begin{equation}\label{Cayley11}
\langle F \rangle_{g}\simeq \frac{2m-2}{m^2-2m}\frac{(N_g)^2}{N_g-1} \sim N_g\,.
\end{equation}
Thus, in the limit of large network size $N_g$, the ATT increases
linearly with the system size.

\subsection{Average trapping time for Vicsek fractals}

Since the above method for computing ATT in $C_{m,g}$ is not applicable to  $V_{f,g}$,
we  use another method to determine $\langle F \rangle_g$ for $V_{f,g}$, which is very different from that
used for $C_{m,g}$.

\subsubsection{Mean first-passage time between two adjacent nodes in a general tree}

In order to determine $\langle F \rangle_{g}$ for $V_{f,g}$,  we first derive a universal formula for MFPT
from one node to one of its neighbors in a general tree.
For a  connected tree, let $e=(u,v)$ denote the edge
in the tree connecting  nodes $u$ and $v$. Evidently, if the edge $(u,v)$ is deleted,  the tree will be
divided into two
subtrees: one   contains node $u$, while the other includes node
$v$. Let $N_{u \leftarrow v}$ denote the number of nodes in the subtree
including node $u$, which is exactly the number of nodes in
the original tree lying closer to $u$ than to $v$, including $u$
itself. Let $F_{uv}$ denote the MFPT for a random walker, starting  from $u$ to reach $v$ for the first
time. Then, $F_{uv}$ can be expressed  in terms of $N_{u \leftarrow v}$ as
\begin{equation}\label{A1}
F_{uv}=2N_{u \leftarrow v}-1.
\end{equation}

Equation~(\ref{A1}) can be readily proved as follows. We consider the tree as a
rooted one with node $v$ being its root. Then, $v$ is the father of
$u$ and $N_{u \leftarrow v}$ is  the number of nodes in the subtree
with root  $u$. By definition,  $F_{uv}$ obeys the following relation:
\begin{equation}\label{A2}
F_{uv}=\frac{1}{k}+\frac{k-1}{k}(R_{uu}+F_{uv})\,,
\end{equation}
where $k$ is the degree of node $u$, and $R_{uu}$ is the average return time for node $u$ in the subtree with
$u$ being its root, defined as the mean time for a walker starting from node $u$ to first  return back
without  visiting node $v$.

The first term on the right hand-side  of Eq.~(\ref{A2})
explains the case that the walker, starting  from node $u$, jumps
directly to the neighboring node $v$ in one single step with probability
$\frac{1}{k}$. And the second term accounts for another case
that the walker first reaches one of the other $k-1$ neighbors  of node $u$ and returns back to $u$ taking
time $R_{uu}$, and then  takes $F_{uv}$ more steps to first hit the target
node $v$. According to the Kac formula~\cite{CoBeMo07,SaDoMe08}, one can easily derive that $R_{uu}=2(N_{u
\leftarrow v}-1)/(k-1)$. Then,
Eq.~(\ref{A2}) can be recast as
\begin{equation}\label{A3}
F_{uv}=\frac{1}{k}+\frac{k-1}{k}\left[\frac{2(N_{u \leftarrow v}-1)}{k-1}+F_{uv}\right]\,,
\end{equation}
from which Eq.~(\ref{A1}) is produced. Equation~(\ref{A1}) is a basic characteristic for random walks on a
tree and is useful for the following derivation of the
key quantity $\langle F \rangle_{g}$ for $V_{f,g}$.

%%%%%%%%%%%%%%%%%%%%%%%%%%%%%%%%%%%%%%%%%%%%%%%%%%%%%%%%%
% Figure  3
%%%%%%%%%%%%%%%%%%%%%%%%%%%%%%%%%%%%%%%%%%%%%%%%%%%%%%%%%%
\begin{figure}
\begin{center}
\includegraphics[width=0.65\linewidth,trim=0 0 0 0]{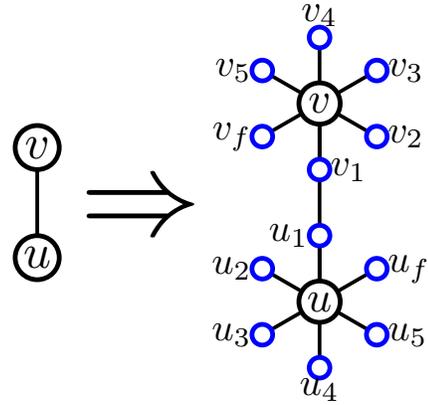}
\end{center}
\caption[kurzform]{(Color online) Second construction method
of the Vicske fractals. $u$ and $v$ are two adjacent nodes in  $V_{f,g-1}$. In generation $g$, each of them
generates $f$ new neighbors, denoted by $u_1$, $u_2$,$\ldots$, $u_f$, and $v_1$, $v_2$,$\ldots$, $v_f$,
respectively. Since $u$ and $v$ are directly connected in  $V_{f,g-1}$, two of their new neighbors (e.g.,
$u_1$ and $v_1$) are linked to each other by a new edge. }\label{Const2}
\end{figure}
%%%%%%%%%%%%%%%%%%%%%%%%%%%%%%%%%%%%%%%%%%%%%%%%%%%%%%%%%%

\subsubsection{An alternative construction of  Vicsek fractals}

Before determining ATT for Vicsek fractals, we introduce another construction approach for this fractal
family. Suppose that we have $V_{f,g-1}$. Then, $V_{f,g}$ can be obtained from $V_{f,g-1}$ as follows (see~Fig.~\ref{Const2}). First, for each node in  $V_{f,g-1}$, $f$ new nodes are created and connected to the
old node. Then, for each pair of adjacent nodes, $u$ and $v$ in $V_{f,g-1}$, a new edge is added between two of
their  new neighboring nodes.
Note that each new node generated in generation $g$ has at most one new neighbor.

In the sequel,  we classify the nodes in $V_{f,g}$  in the following way. We  represent the set of
nodes in $V_{f,g-1}$ as $\Lambda_{g-1}$, and denote the set
of those nodes created in generation $g$ by $\bar{\Lambda}_{g}$.
Obviously, $\Lambda_{g}=\Lambda_{g-1}+\bar{\Lambda}_{g}$.
Moreover, $\bar{\Lambda}_{g}$ can be  separated into two subsets, $\bar{\Lambda}_{g}^{(1)}$ and
$\bar{\Lambda}_{g}^{(2)}$, i.e., $\bar{\Lambda}_{g}=\bar{\Lambda}_{g}^{(1)}\cup \bar{\Lambda}_{g}^{(2)}$,
where $\bar{\Lambda}_{g}^{(1)}$
is the set of nodes with degree 1 and $\bar{\Lambda}_{g}^{(2)}$ is
that with degree 2. Evidently, the cardinality (i.e., the number of nodes) of $\bar{\Lambda}_{g}^{(2)}$ is
$|\bar{\Lambda}_{g}^{(2)}|=2E_{g-1}=2(f+1)^{g-1}-2$, and that of
$\bar{\Lambda}_{g}^{(1)}$ is  $|\bar{\Lambda}_{g}^{(1)}|=f\,N_g-2E_{g-1}=(f-2)(f+1)^{g-1}+2$ .

We next introduce a new quantity for  $V_{f,g}$, i.e., distance of the trap---central node 1, denoted by $D_g$,
defined by
\begin{equation}\label{Distance}
D_{g}=\sum_{i=2}^{N_g} d_{i}(g)\,,
\end{equation}
where $d_{i}(g)$ is the length of the shortest path  from node $i$ to the central node in $V_{f,g}$. According
to the second construction, one has
\begin{eqnarray}\label{C15}
D_{g}&=&\sum_{i\in \bar{\Lambda}_{g}^{\rm (1)}}d_{i}(g)+\sum_{i\in
\bar{\Lambda}_{g}^{\rm (2)}}d_{i}(g)+\sum_{i\in
\Lambda_{g-1}}d_{i}(g)\nonumber\\
%&=&\sum_{i\in \bar{\Lambda}_{g}^{\rm 1}}[1+d_{u_{i}}(g)]+\sum_{i\in
%\bar{\Lambda}_{g}^{\rm 2}}d_{u_{i}}(g)+\sum_{i\in
%\Lambda_{g-1}}d_{i}(g)\nonumber\\
&=&|\bar{\Lambda}_{g}^{\rm (1)}|+(f+1)\sum_{i\in
\Lambda_{g-1}}d_{i}(g)\nonumber\\
&=&(f-2)(f+1)^{g-1}+2+3(f+1)\sum_{i\in
\Lambda_{g-1}}d_{i}(g-1)\nonumber\\
&=&(f-2)(f+1)^{g-1}+2+3(f+1)D_{g-1}\,,
\end{eqnarray}
where the evident relation $d_{i}(g)=3d_{i}(g-1)$ was used.
With initial condition $D_{1}=f$, Eq.~(\ref{C15}) can be solved to yield
\begin{equation}\label{C16}
D_{g}=\frac{(f+1)^{g-1}\left (f^23^{g+1}-3f^2+4f+4 \right)-4}{6f+4}\,,
\end{equation}
which is useful  for the following computation.

\subsubsection{Evolution law for mean first-passage time and trapping time in Vicsek fractals}

Let $F_{ij}(g)$ denote the MFPT of random walks on $V_{f,g}$, staring
from node $i$, to arrive at node $j$ for the first time. And let $(u,v)$ denote an edge
connecting two nodes $u$ and $v$ in $V_{f,g-1}$.

Below, we will derive a
relation governing $F_{uv}(g)$ and $F_{uv}(g-1)$, based on which we will show how the trapping time $F_{i}(g)$
evolves with $g$. For this purpose, we consider $V_{f,g-1}$ as a rooted tree with node $v$ being the root, and
thus
$v$ is the father of $u$. We assume that in the evolution of the
Vicsek fractals, node $v$ is always the root. In addition,
for the rooted fractal family $V_{f,g}$, we use $C_{u}(g)$ to represent the number of nodes in the subtree,
whose root is  $u$. Using the second construction method, it is easy to obtain
\begin{equation}\label{B1}
C_{u}(g)=(f+1)C_{u}(g-1)-1.
\end{equation}

We now begin to derive the relation between $F_{uv}(g)$ and
$F_{uv}(g-1)$. According to the general result given in
Eq.~(\ref{A1}), we have
\begin{equation}\label{B4}
F_{uv}(g-1)=2C_{u}(g-1)-1\,.
\end{equation}
Figure~\ref{Const2} shows that in generation $g$, node $u_1$ is the father of $u$, and node $v$ becomes  an
ancestor of $u$, instead of being $u$'s  father. In addition, $v_1$ (a child of  $v$) is also an ancestor of
$u$. Thus, for a random walker in
$V_{f,g}$, if it wants to transfer from $u$ to $v$, it must pass
through node  $u_1$ and  $v_1$. Therefore,
\begin{eqnarray}\label{B5}
F_{uv}(g)&=&F_{uu_1}(g)+F_{u_1v_1}(g)+F_{v_1v}(g)\nonumber \\
&=&2[C_{u}(g)+C_{u_1}(g)+C_{v_1}(g)]-3\nonumber\\
&=&6C_{u}(g)+3.
\end{eqnarray}
Combining Eqs.~(\ref{B1})-(\ref{B5}), we
obtain
\begin{eqnarray}\label{B7}
F_{uv}(g)&=&6[C_{u}(g)+1]-3\nonumber\\
&=&6(f+1)C_{u}(g-1)-3\nonumber\\
&=&3(f+1)F_{uv}(g-1)+3(f+1)-3\nonumber\\
&=&3(f+1)F_{uv}(g-1)+3f\,.
\end{eqnarray}

We proceed to derive the relation governing $F_{i}(g)$ and
$F_{i}(g-1)$.  For node $i$ in $V_{f,g-1}$, the shortest-path from $i$ to the central node $1$ is unique,
which has  length $d_{i}(g-1)$ and is denoted by  $i=i_{0}\rightarrow i_1 \rightarrow i_2 \rightarrow
i_3\rightarrow \cdots \rightarrow i_{d_{i}(g-1)}=1$. Since the Vicsek fractals  have a treelike structure, we
have
\begin{equation}\label{B8}
F_{i}(g-1)=\sum_{r=1}^{d_{i}(g-1)}F_{i_{r-1}i_{r}}(g-1)\,.
\end{equation}
Equations~(\ref{B7}) and ~(\ref{B8}) give rise to
\begin{eqnarray}\label{B9}
F_{i}(g)&=&\sum_{r=1}^{d_{i}(g-1)}F_{i_{r-1}i_{r}}(g)\nonumber\\
&=&\sum_{r=1}^{d_{i}(g-1)}[3(f+1)F_{i_{r-1}i_{r}}(g-1)+3f]\nonumber\\
&=&3(f+1)\sum_{r=1}^{d_{i}(g-1)}F_{i_{r-1}i_{r}}(g-1)+3fd_{i}(g-1)\nonumber\\
&=&3(f+1)F_{i}(g-1)+3fd_{i}(g-1)\,,
\end{eqnarray}
which is a basic relation between $F_{i}(g)$ and
$F_{i}(g-1)$ and is useful in determining of  ATT  $\langle
F\rangle_{g}$ later on.

\subsection{Closed-form solution to average trapping  time for Vicsek fractals}

Having obtained the intermediate quantities, we are now in a position to determine the ATT  $\langle
F\rangle_{g}$ for $V_{f,g}$. We define the following two
quantities for $n \leq g$:
\begin{equation}\label{C1}
F_{n}^{\rm tot}(g)=\sum_{i\in \Lambda_{n}}F_{i}(g)\,
\end{equation}
and
\begin{equation}\label{C2}
\bar{F}_{n}^{\rm tot}(g)=\sum_{i\in \bar{\Lambda}_{n}}F_{i}(g)\,,
\end{equation}
so,
\begin{equation}\label{C5}
F_{g}^{\rm tot}(g)=F_{g-1}^{\rm tot}(g)+\bar{F}_{g}^{\rm tot}(g)\,.
\end{equation}
Thus, the problem of determining $\langle
F\rangle_{g}$ is reduced to finding $F_{g-1}^{\rm tot}(g)$ and $\bar{F}_{g}^{\rm tot}(g)$. According to
Eq.~(\ref{B9}), we have
\begin{eqnarray}\label{C3}
F_{g-1}^{\rm tot}(g)&=&\sum_{i\in \Lambda_{g-1}}F_{i}(g)\nonumber\\
&=&\sum_{i\in \Lambda_{g-1}}[3(f+1)F_{i}(g-1)+3f\,d_{i}(g-1)]\nonumber\\
&=&3(f+1)F_{g-1}^{\rm tot}(g-1)+3f\,D_{g-1}\,.
\end{eqnarray}
Hence, to obtain $F_{g}^{\rm tot}(g)$, we only need to determine the
quantity $\bar{F}_{g}^{\rm tot}(g)$.

By definition,  $\bar{F}_{g}^{\rm tot}(g)$ can be rewritten as
\begin{equation}\label{C6}
\bar{F}_{g}^{\rm tot}(g)=\sum_{i\in \bar{\Lambda}_{g}^{(1)}}F_{i}(g)+\sum_{i\in
\bar{\Lambda}_{g}^{(2)}}F_{i}(g)\,.
\end{equation}
We begin by determining the first summation term $\sum_{i\in
\bar{\Lambda}_{g}^{(1)}}F_{i}(g)$ on the right hand side of Eq.~(\ref{C6}). For any node $i \in
\bar{\Lambda}_{g}^{(1)}$, it has only one neighbor, i.e., its father node $m_{i} \in \Lambda_{g-1}$. So,
\begin{equation}\label{C7}
F_{i}(g)=1+F_{m_{i}}(g)\,.
\end{equation}
Consequently,
\begin{eqnarray}\label{C8}
\sum_{i\in \bar{\Lambda}_{g}^{(1)}}F_{i}(g)&=&\sum_{i\in
\bar{\Lambda}_{g}^{(1)}}[1+F_{m_{i}}(g)]\nonumber\\
&=&|\bar{\Lambda}_{g}^{(1)}|+\sum_{i\in \bar{\Lambda}_{g}^{(1)}}F_{m_{i}}(g)\nonumber\\
&=&|\bar{\Lambda}_{g}^{\rm
(1)}|+\sum_{i\in\Lambda_{g-1}}F_{i}(g)h_{i}^{(1)}(g)\,,
\end{eqnarray}
where $h_{i}^{(1)}(g)$ is the number of nodes in $V_{f,g}$, which have degree 1
and are adjacent to node $i$ belonging to $V_{f,g-1}$.

We proceed to evaluate the second summation term $\sum_{i\in
\bar{\Lambda}_{g}^{(2)}}F_{i}(g)$ in Eq.~(\ref{C6}). According to
the second construction method of Vicsek fractals discussed above, nodes in $\bar{\Lambda}_{g}^{(2)}$ are
generated in pairs (see~Fig.~\ref{Const2}). For  each edge connecting a pair of adjacent nodes in $V_{f,g-1}$,
e.g., $u$ and $v$ in  $\Lambda_{g-1}$,  there must exist two new nodes, e.g.,  $u_1$ and $v_1$ in
$\bar{\Lambda}_{g}^{(2)}$, satisfying
\begin{equation}\label{C9}
F_{u_1}(g)=\frac{1+F_{u}(g)}{2}+\frac{1+F_{v_1}(g)}{2}
\end{equation}
and
\begin{equation}\label{C10}
F_{v_1}(g)=\frac{1+F_{v}(g)}{2}+\frac{1+F_{u_1}(g)}{2}\,.
\end{equation}
These lead to
\begin{equation}\label{C11}
F_{u_1}(g)+F_{v_1}(g)=4+F_{u}(g)+F_{v}(g)\,
\end{equation}
and
\begin{equation}\label{C12}
\sum_{i\in \bar{\Lambda}_{g}^{(2)}}F_{i}(g)=4E_{g-1}+\sum_{i\in\Lambda_{g-1}}F_{i}(g)h_{i}^{(2)}(g)\,,
\end{equation}
where $h_{i}^{(2)}(g)$ denote the number of nodes in $V_{f,g}$, which have two neighbors with one  being node
$i$  existing in $V_{f,g-1}$ previously.

Notice that for any node $i \in
\Lambda_{g-1}$, we have $h_{i}^{(1)}(g)+h_{i}^{(2)}(g)=f$. Plugging Eqs.~(\ref{C8}) and~(\ref{C12}) into
Eq.~(\ref{C6}) yields
\begin{eqnarray}\label{C13}
\bar{F}_{g}^{\rm tot}(g)&=&|\bar{\Lambda}_{g}^{(
1)}|+\sum_{i\in\Lambda_{g-1}}F_{i}(g)h_{i}^{(1)}(g)\nonumber\\
&&+4E_{g-1}+\sum_{i\in\Lambda_{g-1}}F_{i}(g)h_{i}^{(2)}(g)\nonumber\\
&=&|\bar{\Lambda}_{g}^{\rm
(1)}|+4E_{g-1}+f\, \sum_{i\in\Lambda_{g-1}}F_{i}(g)\nonumber\\
&=&|\bar{\Lambda}_{g}^{\rm (1)}|+4E_{g-1}+f\,F_{g-1}^{\rm tot}(g)\,.
\end{eqnarray}

Substituting Eqs.~(\ref{C3})  and~(\ref{C13}) into Eq.~(\ref{C5}), we
obtain
\begin{eqnarray}\label{C17}
F_{g}^{\rm tot}(g)&=&|\bar{\Lambda}_{g}^{(1)}|+4E_{g-1}+(f+1)F_{g-1}^{\rm tot}(g)\nonumber\\
&=&|\bar{\Lambda}_{g}^{(1)}|+4E_{g-1}+(f+1)\nonumber\\
&&\times[3(f+1)F_{g-1}^{\rm tot}(g-1)+3fD_{g-1}]\,.
\end{eqnarray}
Combining the above-obtained results and using the initial condition $F_{1}^{\rm tot}(1)=f$, one can solve Eq.~(\ref{C17})
 to yield
\begin{eqnarray}\label{C18}
F_{g}^{\rm tot}(g)&=&\frac{1}{6 f+4}[4 f\times3^g (f+1)^{2 g-1}-(f+1)^{g-1}\times\nonumber\\
&& \left( f^2 3^{g+1}-3 f^2+8 f+4\right)+4]\,,
\end{eqnarray}
Plugging the last expression into Eq.~(\ref{ATT}), we arrive at the
explicit formula for the ATT on $V_{f,g}$ as follows:
\begin{eqnarray}\label{C19}
\langle
F\rangle_{g}&=&\frac{1}{N_{g}-1}\sum_{i=2}^{N_{g}}F_{i}(g)=\frac{1}{N_{g}-1}F_{g}^{\rm tot}(g)\nonumber\\
%&=&\frac{1}{2(1+f)(2+3f)\left(-1+(1+f)^g\right)}\nonumber\\
%&&\times [4 f \left(1-2 (1+f)^g+3^g
%\left((1+f)^2\right)^g\right)+4\nonumber\\
%&&-4 (1+f)^g-3 \left(-1+3^g\right) f^2 (1+f)^g]\,.
&=&\frac{1}{(6f+4)\left[(f+1)^g-1\right]}[4f \times3^g  (f+1)^{2 g-1}\nonumber\\
&&-(f+1)^{g-1}\left(f^2 3^{g+1} -3 f^2+8 f+4\right)+4],\nonumber\
\end{eqnarray}
which can be further expressed as a function of the network size $N_{g}$, as
\begin{eqnarray}\label{C19a}
\langle
F\rangle_{g}&=&\frac{2f}{3f^2+5f+2}\frac{(N_g)^{\frac{\ln 3}{\ln (f+1)}+2}}{N_g-1}\nonumber\\
&&-\frac{3f^2}{6f^2+10f+4}\frac{(N_g)^{\frac{\ln 3}{\ln (f+1)}+1}}{N_g-1}\nonumber\\
&&+\frac{3f^2-8f-4}{6f^2+10f+4}\frac{N_g}{N_g-1}\nonumber\\
&&+\frac{2}{3f+2}\frac{1}{N_g-1}\,.
\end{eqnarray}
Equation~(\ref{C19a}) unveils the succinct dependence relation
of ATT on the network size $N_g$ and parameter $f$.  When $N_{g}\to\infty$, we have  the following  leading term
for $\langle F \rangle_g$:
\begin{eqnarray}\label{C20}
\langle F\rangle_{g}\sim (N_{g})^{1+\ln3/\ln(f+1)}\,.
\end{eqnarray}
Thus, ATT grows approximately as a power-law function in the
network order $N_g$ with the exponent $1+\ln 3/\ln (f+1)$
$>$ 1 and being a decreasing function of $f$. This is in sharp contrast with that obtained for
Cayley trees, indicating that the underlying topologies play an essential role in trapping efficiency for polymer networks. At the same time, the different scalings obtained reveal some information about the underlying microscopic structures of the polymer networks.

\section{Conclusions}

To explore the effect of the underlying structures on the trapping efficiency, we have studied the trapping problem defined on two families of polymer networks, i.e., Cayley trees and Vicsek
fractals, concentrating on a particular case with the single trap positioned at the central node.  Using two different techniques, we have obtained analytically the closed-form solutions for ATT for both cases, based on which we have further expressed ATT in terms of the network sizes. Our results show that for large systems, the leading behaviors of ATT for Cayley trees and Vicsek fractals follow distinct scalings, with the trapping efficiency of the former being much higher than that of the latter. Our work unveils that the geometry of macromolecules has a substantial influence on the trapping efficiency in polymer networks.

Actually, in addition to the trapping problem, other dynamics for Cayley trees and Vicsek
fractals also display quite different behaviors, e.g., relaxation~\cite{GuBl05} and energy transfer~\cite{BlVoJuKo05}. The reason lies in the fact that the dynamics is, in all three cases, determined by the topological structures of the polymer networks. Although the physical situations and measurements of these three dynamics are distinct, they are related to one another, since they all are encoded in eigenvalues of matrices (transition matrix for trapping problems~\cite{Lo96,ZhWuCh11}, Laplacian matrix for the other two dynamics~\cite{BlVoJuKo05,GuBl05}) of the polymer networks, which reflect the global topological properties of the underlying structures. Thus, our results provide some new insight, allowing an easy differentiation between the structures of the two important classes of polymer, Cayley trees and Vicsek fractals.

\begin{acknowledgments}
This work was supported by the National Natural Science Foundation
of China under Grant No. 61074119 and the Hong Kong Research Grants
Council under the General Research Funds Grant CityU 1114/11E.
\end{acknowledgments}

\nocite{*}
%\bibliography{aipsamp}% Produces the bibliography via BibTeX.

\end{document}